\documentclass{aa}

\usepackage{graphicx}

\usepackage{natbib}
\usepackage{txfonts}

\begin{document}

\title{The Effect of Different Type Ia Supernova Progenitors on Galactic Chemical Evolution} 

\titlerunning{Different Supernova Progenitors}

\author{F. Matteucci\inst{1,2}\thanks{email to: matteucc@oats.inaf.it}
\and E. Spitoni\inst{1}
\and S.Recchi\inst{3}
\and R.Valiante\inst{4}}

\authorrunning{Matteucci \& al.}

\institute{Dipartimento di Astronomia, Universit\'a di Trieste,  Via G. B. Tiepolo 11, I-34143 Trieste, 
Italy 
\and  I.N.A.F. Osservatorio Astronomico di Trieste, Via G. B. Tiepolo 11, I-34143 Trieste (TS), Italy
\and Institute of Astronomy, Vienna University, T\"urkenschanzstrasse 17,
A-1180, Vienna, Austria
\and Dipartimento di Astronomia, Universita di Firenze, Largo E. Fermi 5, 50125 Firenze}

\date{Received xxxx / Accepted xxxx}

\abstract{}
{Our aim is to show how different hypotheses about Type Ia supernova progenitors can affect Galactic chemical evolution. Supernovae Ia, in fact, are believed to be the main producers of Fe and the 
timescale with which Fe is restored into the interstellar medium depends on the assumed supernova progenitor 
model. This is a way of selecting the right progenitor model for supernovae Ia, a still debated issue.} 
{We include different Type Ia SN progenitor models, identified by their distribution of time delays, in a very detailed 
chemical 
evolution model for the Milky Way which follows the evolution of several chemical species. We test the single degenerate 
and the double 
degenerate models for supernova Ia progenitors, as well as other more empirical models based on differences in the time delay 
distributions.}  
{We find that assuming the single degenerate or the double degenerate scenario produces negligible differences in the predicted [O/Fe] vs. [Fe/H] relation. On the other hand, assuming a percentage of prompt (exploding in the first 100 Myr) Type Ia supernovae of 50\%, or that the maximum Type Ia rate is reached after 3-4 Gyr from the beginning of star formation, as suggested by several authors, produces more noticeable effects on the [O/Fe] trend. However, given the spread still existing in the observational data no model can be firmly excluded on the basis of only the [O/Fe] 
ratios. On the other hand, when the predictions of the different models are compared with the G-dwarf metallicity distribution, the scenarios with very few prompt Type Ia supernovae can be excluded.}
{Models including the single degenerate or double degenerate scenario with a percentage of 10-13\% of prompt Type Ia supernovae produce results in very good agreement with the observations.  A fraction of prompt Type Ia supernovae larger than 30\% worsens the agreement with observations and the same occurs if no prompt Type Ia supernovae are allowed. In particular, two empirical models for the Type Ia SN progenitors can be excluded:
the one without prompt Type Ia supernovae and the one assuming delay time distribution going like $\propto t^{-0.5}$. We conclude that the typical timescale for the Fe enrichment in the Milky Way is around 1-1.5 Gyr and that Type Ia supernovae should appear already during the halo phase.}


\maketitle

\section{Introduction}
The Type Ia supernova (SN) rate, which depends on the SN Ia progenitor model and the star formation history, is a 
fundamental ingredient in models of galactic chemical evolution. 
The progenitor model can be described simply by a delay time distribution (DTD) 
function, which is the distribution of the explosion times.
In the pioneering work of Greggio \& Renzini (1983a, hereafter GR83) 
there was, for the first time, an expression for the Type Ia SN rate in the scenario of the single degenerate model. 
In this scenario, SNe Type Ia arise from the explosion of a C-O white dwarf in 
a close binary system where the companion is either a red giant or a main 
sequence star (Whelan \& Iben, 1973; Munari \& Renzini, 1992; 
Kenyon et al.1993; Hachisu et al., 1996; 1999; Han \& Podsiadlowski, 2004). The DTD of this model arises from the above 
described scenario and predicts the explosion times of binary systems with the appropriate characteristics.
An alternative model for progenitors of Type Ia SNe was proposed by 
Iben \& Tutukov (1984). In this scenario, two C-O white dwarfs merge after loss of angular momentun due to gravitational wave 
emission, and explode because the final object reaches the Chandrasekhar mass. 
Tornamb\'e \& Matteucci 
(1986) formulated a Type Ia SN rate in this scenario and applied it to 
galactic chemical evolution models.
In the following, Matteucci \& Greggio (1986) tested the GR83 rate by means of a detailed 
chemical 
evolution model of the Milky Way and interpreted the [$\alpha$/Fe] ratios versus [Fe/H] as due to the 
delay in the Fe production from Type Ia SNe, thus confirming previous suggestions (Tinsley, 1979; 
Greggio \& Renzini 1983b). 
In recent years other authors (Dahlen et al. 2004; Strolger et al. 2004; Mannucci et al. 2005, 2006; Scannapieco \& 
Bildsten, 2005; Sullivan et al. 2006; Aubourg  et al. 2008; Pritchet et al. 2008; Totani et al. 2008) have proposed 
Type Ia SN rates based on DTDs derived 
empirically. One example is the rate of Mannucci et al. (2005;2006): they suggested that two 
populations of progenitors of Type Ia SNe are needed to explain the dependence of the rates 
on the colors of the parent galaxy. The presence of Type Ia SNe in old, red, 
quiescent galaxies is an indication that part of these SNe originate from old 
stellar populations. On the contrary, the increase of the rate in blue galaxies
(by a factor of 30 going from (B-K) $\sim$2 to (B-K) $\sim$ 4.5) shows that 
part of the SN Ia progenitors is related to young stars and closely follows 
the evolution of the  star formation rate (hereafter SFR).
Mannucci, Della Valle \& Panagia (2006, hereafter MVP06), on the 
basis of the previously described relation between the Type Ia SN rate and the color of the parent 
galaxies, their radio power as measured by Della Valle et al. (2005), and cosmic age, 
concluded that there are two populations of progenitors of 
Type Ia SNe. They have suggested (but see Greggio, Renzini \& Daddi, 2008) that the current observations can be 
accounted for only if about half of the SNe Ia ({\it prompt SNe Ia}) explode within $10^{8}$ years after the 
formation of their progenitors, while the rest explode during a wide period of time extending up to 
10 Gyrs ({\it tardy SNe Ia}).
Matteucci et al. (2006) applied this formulation of the Type Ia SN rate to chemical evolution 
models of galaxies of different morphological type. They concluded that a fraction of 50\% of prompt 
Type Ia SNe worsens the agreement with abundance data, especially in the Milky Way, and 
suggested that prompt Type Ia SNe should indeed exist but their fraction should not be larger 
than 30\%. 
They also found a possible scenario for SN Ia progenitors in the MVP06 framework: a 
possible justification for this strongly bimodal DTD can be found in the framework of the single 
degenerate model for the progenitors of Type Ia SNe. In fact, such a DTD can be found if one 
assumes that the function describing the distribution of mass ratios inside the binary systems, 
is a multi-slope function. In  particular, this choice means that in the range 5-8 $M_{\odot}$ are 
preferred the systems where $M_1 \sim M_2$, whereas for lower mass progenitors are 
favored systems where $M_1 >> M_2$. 
Another suggestion came from Dahlen et al. (2004) and Strolger et al.(2004, 2005), on the basis of the observational 
cosmic 
Type Ia SN rate. They suggested a DTD with no prompt Type Ia SNe with a maximum occurring at 
3-4Gyr. This DTD has never been tested in chemical evolution models. 
More recently, a direct determination of the DTD function has been reported by Totani et al. (2008), on the basis of the faint variable objects detected in Subaru/XMM-Newton Deep Survey (SXDS). They concluded that the DTD function is inversely proportional to the delay time, i.e. the DTD can be well described by a featurless power law (DTD $\propto t^{-n}$, with $n \sim 1$). A similar suggestion came from Pritchet et al. (2008), who suggested a  DTD $ \propto t^{-0.5 \pm 0.2}$.
  
The maximum at which the the Type Ia SN rate occurs  is a very important parameter which affects the shape of the [$\alpha$/Fe] relation in galaxies.
This timescale  depends on the fraction of prompt/tardy Type Ia SNe and on the specific history of 
star formation in galaxies, as shown by Matteucci \& Recchi (2001), being shorter than in the solar neighbourhood for ellipticals and 
longer for irregular systems.

In this paper, for the first time we show the effects of adopting different DTDs in a model for the chemical evolution of the Milky Way: in particular, we will focuse on three relevant observational costraints: i) the [O/Fe] vs.[Fe/H] relation in the solar vicinity for which many accurate data are now available, ii) the Type Ia SN rate at the present time and iii) the G-dwarf metallicity distribution. To do that we we adopt five different DTDs, two of which are empirically derived.

The paper is organized as follows:
in Section 2 the formulation for the SNIa rates is presented.
In Section 3 the galactic chemical 
evolution model is described and the results are discussed, and in Section 4 some conclusions are 
drawn.

\section{Type Ia supernovae}
\subsection{Progenitors}

We recall here the most common models for the progenitors of Type Ia
SNe proposed insofar:

\begin{itemize}

\item The merging of two C-O white dwarfs (WDs), due to gravitational wave radiation,
which reach the Chandrasekhar mass ($\sim 1.4
M_{\odot}$) and explode by C-deflagration (Iben
and Tutukov 1984). This is known as double-degenerate (DD)
scenario. In the original paper of Iben \& Tutukov the progenitor masses are in the range 5-9$M_{\odot}$ to ensure 
two WDs of at least $\sim 0.7 M_{\odot}$ in order to reach the Chandrasekhar 
mass. 
 In Greggio (2005, hereafter G05) the range of progenitors is 2.0-8.0$M_{\odot}$, so that different combinations of WD masses are allowed, although always 
leading to exceed the  Chandrasekhar mass.
The clock for the explosion in this scenario 
is given by the lifetime of the secondary plus the gravitational time delay. 

\item The C-deflagration of a Chandrasekhar mass C-O WD after accretion from a non-degenerate companion
(Whelan and Iben 1973; Munari and Renzini 1992; Kenyon et
al. 1993). This model is known as the single-degenerate (SD) one. The
main problem with this scenario is the narrow range of permitted
values of the mass accretion rate in order to obtain a stable
accretion, instead of an unstable accretion with a consequent nova
explosion and mass loss. In this case, in fact, the WD never achieves
the Chandrasekhar mass. In particular, Nomoto, Thielemann \& Yokoi
(1984) found that a central carbon-deflagration of a WD results for a
high accretion rate ($\dot M > 4 \cdot 10^{-8}\, M_{\odot} \,{\rm
yr}^{-1}$) from the secondary to the primary star (the WD). They found
that $\sim (0.6 - 0.7)\; M_{\odot}$ of Fe plus traces of elements from C
to Si are produced in the deflagration, well reproducing the observed
spectra. The clock for the explosion here is given by the lifetime of the secondary star.

\item A sub-Chandrasekhar C-O WD exploding by He-detonation induced by
accretion of He-rich material from a He star companion (Tornamb\'e and Matteucci 1987;
Limongi and Tornamb\'e 1991).

\item A further model by Hachisu et al. (1996; 1999) is based on the
classical scenario of Whelan and Iben (1973), but they find an important
metallicity effect. When the accretion process begins, the primary
star (WD) develops an optically thick wind which helps in stabilizing
the mass transfer process. When the metallicity is low ($[Fe/H] <
-1$), the stellar wind is too weak and the explosion cannot
occur. The clock for the explosion here is also given by the lifetime of the secondary 
star. However, one has to take into account that Type Ia SN progenitor systems do not form before the gas has attained a threshold metallicity. Therefore, the chemical enrichment from Type Ia SNe is delayed because of this effect.

\end{itemize}

\subsection{The assumed DTDs}

We have tested the effects of five DTDs:
\begin{itemize}
\item The Matteucci \& Recchi (2001, hereafter MR01) one. which is practically the same as GR83, and is based on the single degenerate model for Type I SN progenitors (see Matteucci et al. 2006 for details). In this DTD the fraction of prompt Type Ia SNe is 13\%.This formulation has proven one of the best to describe the chemical evolution of most galaxies. The very first SNe Ia to explode in this scenario are systems made of an 8+8 $M_{\odot}$ which takes $\sim 30-35$ Myr. 

\item The DTD suggested by G05 for the DD scenario: in particular, the DTD characterized by the exponent of the distribution function of final separations $\beta_a=-0.9$, by the exponent of the gravitational delay time distribution $\beta_g=-0.75$ and by a maximum nuclear delay time (lifetime of the secondary star) $\tau_{n,x}=0.4$ Gyr. The reason for chosing these parameters resides in the fact that they ensure a not too flat distribution of separations, which seems to be required to reproduce the specific Type Ia SN rate in galaxies as a function of colors (see Greggio and Cappellaro 2009). In this DTD the fraction of prompt Type Ia SNe is 10\%. The very first SNe Ia to explode in this scenario are systems made of an 8+8 $M_{\odot}$ which takes $\sim 30-35$ Myr plus the minimum gravitational time delay (1 Myr, G05).

\item The DTD of MVP06.  Here the fraction of prompt Type Ia SNe is 50\%. The very first systems explode also after 30-35 Myr.

\item The DTD proposed by Strolger et al. (2004, hereafter S04) with a maximum at 3-4 Gyr. Here the very first systems explode after $2.5 \cdot 10^{8}$ years. Therefore, there are no prompt Type Ia SNe.

\item The DTD proposed by Pritchet et al. (2008, PHS08), with DTD 
$\propto t^{-0.5}$, where the fraction of prompt Type Ia SNe is $\sim 4\%$.
The one proposed by Totani et al. (2008) goes like $t^{-1}$ and we did not test it since these authors have already shown that it is very similar to the DTD of G05 for the DD scenario.

\end{itemize}
In Figure 1 we show the different DTDs including those mentioned above. As one can see, while the DTDs of the SD and DD scenarios are similar, those of MVP06 and S04 and PHS08 are quite different. We did not take into account the Hachisu et al. (1996; 1999) model since its effects have been already discussed in Kobayashi et al. (1998) and MR01.  We did not consider also the sub-Chandra model because the correponding SN rate was already computed by Tornamb\'e and Matteucci (1987). Finally, we show the DTD relative to the DD model (close channel) but we will not use it for computing galactic chemical evolution, and the reason for this is that it is very similar to the DD (wide channel) DTD.

\begin{figure}
\includegraphics[width=0.5\textwidth]{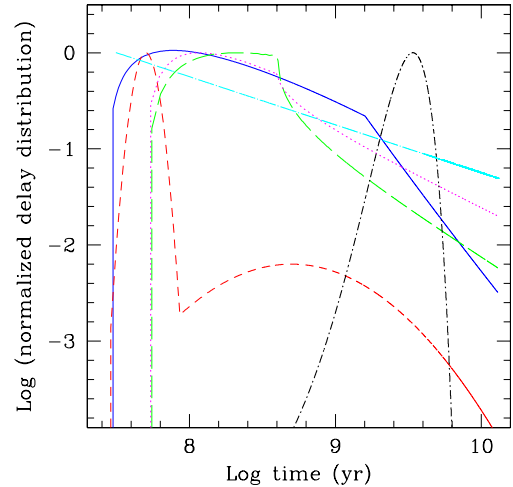}
\caption{Various DTD functions normalized to their own maximum value: the continuous blue line is the DTD of MR01; the long dashed green line is the DTD of G05 for the DD wide channel; the dotted magenta line is the DTD for the DD close channel of G05; the dashed red line is the DTD of MVP06; the short dashed-dotted black line is the DTD of S04, the cyan long dashed-dotted line is the DTD of PHS08. }
\label{fig1}
\end{figure}

\subsection{The calculation of the Type Ia SN rate}
Once a DTD  has been assumed, the Type Ia SN rate is computed according to
Greggio (2005):

\begin{equation}
R_{Ia}(t)=k_{\alpha} \int^{min(t, \tau_x)}_{\tau_i}{A (t-\tau) \psi(t-\tau) 
DTD(\tau) d \tau}
\end{equation}

where $\psi(t)$ is the SFR and $A(t- \tau)$ is the fraction of binary systems which give rise  to Type Ia SNe and in principle it can vary in time. Here we will assume $A$ to be a constant. It is worth noting that the fraction A  represents the fraction of binary systems with those particular characteristics 
to give rise to Type Ia SNe relative to the whole range of star masses (0.1-100$M_{\odot}$).
The time $\tau$ is the delay time defined in the range $(\tau_i, \tau_x)$ so that:

\begin{equation}
\int^{\tau_x}_{\tau_i}{DTD( \tau) d \tau}=1
\end{equation}

where $\tau_i$ is the minimum delay time for the occurrence of Type Ia SNe, 
in other words the time at which the first SNe Ia start occurring, and $\tau_x$ is the maximum delay time. 
Clearly these delay times vary according to the 
assumed progenitor scenario: in particular, in the SD scenario $\tau_x$ is just the maximum nuclear delay time, namely the lifetime of the smallest secondary star, whereas in the DD model is the sum of the maximum nuclear delay time and the maximum gravitational delay time.

Finally, $k_{\alpha}$ is the number of stars per unit mass in a stellar 
generation and contains the IMF.

In particular:
\begin{equation}
k_{\alpha}= \int^{m_U}_{m_L}{\phi(m)dm}
\end{equation}

with the normalization condition for the IMF being:
\begin{equation}
\int^{m_U}_{m_L}{m \phi(m)dm} =1
\end{equation}

where $m_L=0.1 M_{\odot}$ and $m_U=100 M_{\odot}$ and define the whole range 
of existence of the stars.

\section{The model for the Milky Way}

The model we adopt for computing the chemical evolution of the Milky Way is based on the original two-infall model by Chiappini et al (1997; 2001). In particular, they assumed that the first infall episode was responsible for the formation of the halo and thick-disk stars that originated from a fast dissipative collapse. The second infall episode formed the thin-disk component, with a timescale much longer than that of the thick-disk formation. The timescale for the formation of the halo-thick disk is 1-2 Gyr, while the timescale for the thin disk is 7 Gyr in the solar vicinity. The authors included in the model also a threshold in the gas density, below which the star formation process stops. The existence of such a threshold value is suggested by observations relative to the star formation in external disk galaxies (Kennicutt 1998, but see Boissier et al. 2007). The physical reason for a threshold in the star formation is related to the gravitational stability, according to which, below a critical density, the gas is stable against density condensations and, consequently, the star formation is suppressed.\\ 
In the two-infall model the halo-thick disk and the thin disk evolutions occur at different rates and they are independent, mostly as a result of different accretion rates. With these precise prescriptions it is possible to reproduce the majority of the observed properties of the Milky Way and this shows how important is the choice of the accretion law for the gas coupled with the SFR in the Galaxy evolution. Some of the most important observational constraints are represented by the various relations between the abundances of metals (C, N, $\alpha$-elements, iron peak elements) as functions of the [Fe/H] abundance (the most common tracer of metallicity) and by the G-dwarf metallicity distribution. The Galactic disk is approximated by a series of concentric annuli, 2 kpc wide, without exchange of matter between them. In this model the thin disk forms ``inside-out'', in the sense that the timescale for the disk formation increases with galactocentric distance. This choice was dictated by the necessity of reproducing the abundance gradients along the Galactic disk (see Chiappini et al. 2001).

The time-evolution of the fractional mass of the element $i$ 
in the gas within a galaxy, $G_{i}$, is described by the basic 
equation:

\begin{equation}
\dot{G_{i}}=-\psi(t)X_{i}(t) + R_{i}(t) + (\dot{G_{i}})_{inf} -
(\dot{G_{i}})_{out}
\end{equation}

where $G_{i}(t)=M_{g}(t)X_{i}(t)/M_{tot}$ is the gas mass in 
the form of an element $i$ normalized to a total fixed mass 
$M_{tot}$ and $G(t)= M_{g}(t)/M_{tot}$ is the total fractional 
mass of gas present in the galaxy at the time t. For modelling spiral disks the masses are substituted by the surface mass densities.
The quantity $X_{i}(t)=G_{i}(t)/G(t)$ represents the 
abundance by mass of an element $i$, with
the summation over all elements in the gas mixture being equal 
to unity. \\
$R_{i}(t)$ represents the returned fraction of matter in the 
form of an element $i$ that the stars eject into the ISM through 
stellar winds and supernova explosions; this term contains all 
the prescriptions concerning the stellar yields and 
the supernova progenitor models.\\
The assumed SFR ($\psi(t)$)
is a Schmidt (1955) law with a dependence on the surface gas density ($k=1.5$, see Kennicutt 1998) and also on the total surface mass density (see Dopita \& Ryder 1994). In particular, the SFR is based on the law originally suggested by Talbot \& Arnett (1975) and then adapted by Chiosi(1980):

\begin{equation}
\psi(r,t)=\nu\left(\frac{\Sigma(r,t) \Sigma_{gas}(r,t)}{\Sigma(r_{\odot},t)^{2}}\right)^{(k-1)}\Sigma_{gas}(r,t)^{k}
\end{equation}

where $t$ is the time, $r$ is the galactocentric distance and the constant $\nu$ is the efficiency of the star formation process and is expressed in $Gyr^{-1}$: in particular,  $\nu= 2 \,Gyr^{-1}$ for the halo and $1 \, Gyr^{-1}$ for the disk ($t\ge 1\, Gyr$). The total surface mass density is represented by $\Sigma(r,t)$, whereas $\Sigma(r_{\odot},t)$ is the total surface mass density at the solar position, assumed to be $r_{\odot}=8$ kpc (Reid 1993). The quantity $\Sigma_{gas}(r,t)$ represents the surface gas density. These choices of values for the parameters allow the model to fit very well the observational constraints, in particular in the solar vicinity. A threshold gas density for the star formation in the thin disk of $7 M_{\odot} pc^{-2}$ is adopted in all the models presented here.
It is worth noting that the SFR described by eq.(6) is equivalent to a simple Kennicutt's law ($SFR=\nu \Sigma_{gas}^{k}$) for what concerns the evolution of chemical abundances, as shown in Colavitti et al. (2009). We have used here the formulation of eq.(6) for continuity with all our previous papers. The IMF is that of Scalo 
(1986) normalized over a mass range of 0.1-100 $M_{\odot}$ and it is assumed to be constant in space and time.

The nucleosynthesis prescriptions are common to all models and are taken from: 
Woosley \& Weaver (1995) yields 
for massive stars (those of Fe relative to the solar chemical composition and those of O metallicity dependent since 
this is the best combination for reproducing the observations, as shown by Fran\c cois et al. 2004); van den 
Hoeck $\&$ Groenewegen 
(1997) metallicity dependent yields for low and intermediate mass stars 
($0.8 \le M/M_{\odot} \le 8$) and Nomoto et al.'s yields 
(1997) for Type Ia SNe. The choice of these particular yields resides in the fact that they give the best agreement with the observed abundances in the solar vicinity, as shown by Fran\c cois et al. (2004). Moreover, the O yields are those best known, since the majority of nucleosynthesis studies agree on this particular element. The yields of Fe from massive stars are still more uncertain and they vary from author to author: however, most of Fe is assumed to be produced by Type Ia SNe and it is known from observational studies, which suggest that these SNe produce, on average, $0.5M_{\odot}$ of Fe (see Blanc \& Greggio, 2008 and references therein).
The assumed stellar lifetimes for all galaxies are those suggested by 
Padovani \& Matteucci (1993).

The two terms 
$(\dot{G_{i}})_{inf}$ and  $(\dot{G_{i}})_{out}$ account for 
the infall of external gas and for galactic winds, respectively.
For the Milky Way we assume no wind, therefore  $(\dot{G_{i}})_{out}=0.$.\\
The infall instead is an important physical process in the Galaxy.
We assume the two-infall law of Chiappini et al. (1997):

\begin{equation}
A(r,t)= a(r) e^{-t/ \tau_{H}(r)}+ b(r) e^{-(t-t_{max})/ \tau_{D}(r)}
\end{equation}

where $a(r)$ and $b(r)$ are two parameters fixed by reproducing the total 
present time surface mass density along the Galactic disk. 
It is worth noting that this particular infall law has proven to be consistent 
with an accretion law derived in a 
cosmological context, by assuming that gas accretion follows the accretion of the dark matter halo for a galaxy  
like the Milky Way (Colavitti et al. 2008). Colavitti et al. (2008) tested various infall laws including cosmologically derived ones and costant and linear laws. These different infall laws influence the SFR which depends on the amount of gas present in the galaxy at any time and therefore the chemical abundances and the stellar metallicity distribution. They concluded that the best law in order to fit the main properties of the Milky Way is still eq. (7) which has also
been adopted in other chemical evolution studies (e.g. Chang et al. 1999; Alib\`es et al. 2001).

In the solar vicinity the total surface mass density $\sigma_{tot}$= 51 $\pm$ 6 $M_{\odot} \; pc^{-2}$  
(see Boissier \& Prantzos 1999). $t_{max}=1.0$ Gyr is the time for the maximum infall on the 
thin  disk, $\tau_{H}= 2.0$ Gyr
is the time scale for the formation of the halo thick-disk and $\tau_{D}(r)$
is the timescale for the formation of the thin disk and it is a function of the galactocentric 
distance (formation inside-out, Matteucci and Fran\c cois 1989;
Chiappini et al. 2001).

In Figure 2 we show the star formation rate as predicted by eq. (6) in the 
framework of the  two-infall model. Note 
the gap in the star formation rate at about  1Gyr; this corresponds to the 
transition phase between the halo-thick 
disk  and the thin disk phase. The gap is due to the fact that the disk is 
formed by means of the second infall 
episode and star formation can occur only if the gas density is above the 
threshold. Therefore, the gap occurs as a 
natural consequence of the two infall episides and a threshold gas density 
for star formation.
It is worth noting that our chemical results would not be much affected 
if we had considered only 
one infall episode and no threshold in the gas density. In this case, the SFR 
would be simply 
exponential without oscillations and there would be no gap between the thick 
and thin disk.
The [O/Fe] vs. [Fe/H] (see next section) would differ only for the lack of a 
loop in the 
transition phase between thick and thin disk (see for example Matteucci \& 
Fran\c cois 1989;1992), as it is instead predicted by the two-infall model.

\begin{figure}
\includegraphics[width=0.5\textwidth]{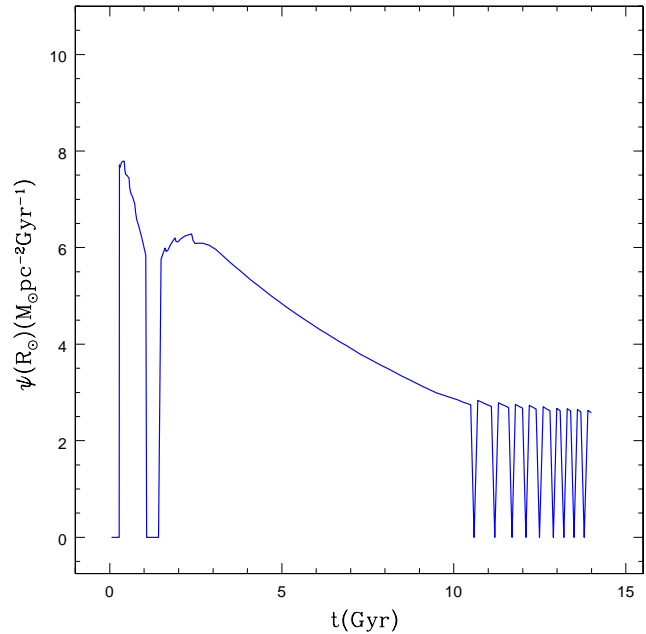}
\caption{The star formation rate as predicted by the two-infall model in the solar neighbourhood: 
the first peak represents the star formation rate during the halo-thick disk phase. Note that there 
is a short gap in the star 
formation between the end of the thick disk phase and the beginning of the thin disk phase: this 
is due to the adoption of a threshold density for 
the star formation. Note that the oscillating behaviour in the last few Gyr is due to the fact that 
the threshold density is easily reached al late times and star formation stops for a while until the gas restored by dying stars increases again the interstellar gas density.}
\label{fig2}
\end{figure}

\section{Predicted [O/Fe] vs. [Fe/H] with differents DTDs}

In Figure 3 we show the predicted Type Ia SN rates for the solar vicinity 
in the framework of the two-infall model previously described and for the considered DTDs. We have chosen the parameter $A$ in eq. (1) in order to reproduce the present time TypeI a SN rate in the Galaxy ($\sim 0.3\, SNe\, century^{-1}$, 
Cappellaro et al. 1999). It is worth noting that a value of A$\sim$ 0.0025 for all the DTDs was chosen, in agreement with Matteucci et al. (2006), with the exception of the PHS08 DTD for which the parameter A=0.0002.
The PHS08 DTD is, in fact, predicting the largest fraction of SNe
Ia exploding with large delays (see Fig. 1). That means that a larger
fraction of long-living progenitors (born several Gyr ago, when the SFR
was much higher, see Fig. 2), relative to the other models, is exploding at the present time.
Therefore, a smaller value of A is required to reproduce the present-time
SN Ia rate in the Galaxy.
The oscillating behaviour shown by some Type Ia SN rates in the last couple of Gyr is clearly related to the oscillations in the star formation rate. In particular, models with a larger percentage of prompt Type Ia SNe show the oscillations at late times
(SD model and MVP06 model), whereas the models of S04, G05 and PHS08, having null, 10\% and 4\% fractions of prompt Type Ia SNe, respectively, show negligible oscillations. 
\begin{figure}
\includegraphics[width=0.5\textwidth]{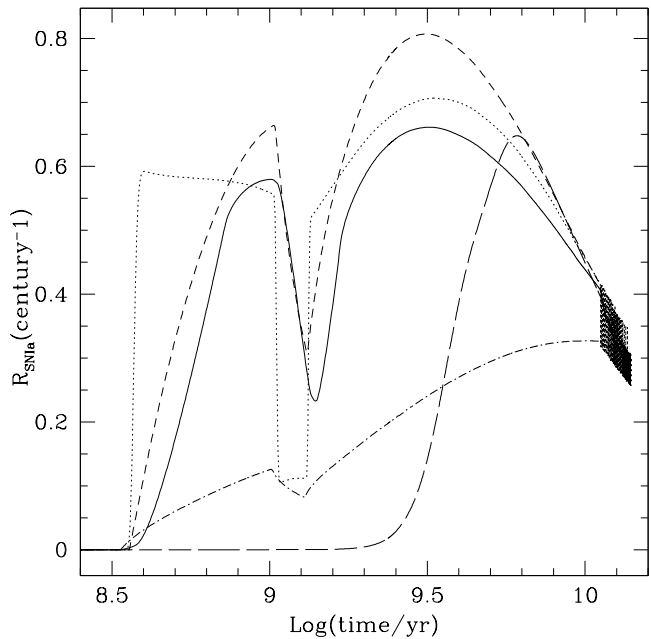}
\caption{The predicted Type Ia SN rates in the solar vicinity in the framework of the 
two-infall model. The five curves refer to different DTDs. In particular, the dotted line represents the DTD of MVP06; the continuous line represents the SD model with the DTD from MR01 and the short dashed line the DD model with the DTD of G05 for the DD wide channel. Finally, the long dashed line represents the DTD suggested by S04 and the dashed-dotted line the DTD 
of PHS08. The rates are expressed in SN per century, and are all normalized to reproduce the present time Type Ia SN rate 
in the Milky Way.}
\label{fig3}
\end{figure}

As one can see, it is evident the effect of the gap in the star formation rate,  
as predicted by the two-infall model,
for the DTDs of MVP06, SD, DD and PHS08 models. For the DTD of S04,  
there is no gap  since the minimum delay in the occurrence of Type Ia SNe is 
much longer than for the other DTDs. There are no prompt Type Ia SNe and 
most of the SNe Ia start occurring only after the gap in the star formation.
In Figure 4, 5 and 6 we show the effect of different DTDs on the [O/Fe] vs. [Fe/H] diagram.
As it is well known this diagram is interpreted on the basis of the ``time-delay model''. 
The time delay model suggests that the [O/Fe] in the halo stars ([Fe/H]$<$ -1.0) is slowly varying with [Fe/H] 
and reflects mainly the behaviour of the O/Fe ratio in massive stars: therefore, one expects a slight 
variation with the stellar mass. 
For [Fe/H]$>$ -1.0, the [O/Fe] ratio starts declining as a consequence of the increase of the Fe production from Type Ia SNe. Clearly, the DTD is fundamental in determining the moment, and therefore the [Fe/H] value, at which this process 
becomes important. We remind that this effect does not depend on details of the star formation history but mainly on the time delay between the chemical 
enrichment from SNe II and SNe Ia.
On the basis of what just said, it is therefore important to see which 
DTD best fits the abundance data. In Figure 4 we show the results for the DTDs related to the SD and DD (wide channel) models.
As one can see, the differences between the results obtained by means of the SD and DD models are small, as it was expected by the comparison of the  DTDs connected to these scenarios (see also Valiante et al. 2009).
In particular, the amount of prompt Type Ia SNe in the SD and DD scenarios is very similar, namely
13\% and 10\%, respectively, as opposed to the DTD of MVP06 where the percentage of prompt SNe 
Ia is 50\% . This difference is evident in the [O/Fe] ratios predicted by the MVP06 DTD (Fig.5) which are declining 
before and faster than for the SD and DD models. In fact, in the model with MVP06 DTD the [O/Fe] knee occurs at [Fe/H]$\sim$-1.8 dex, whereas for the results obtained with the SD and DD DTDs the knee occurs at [Fe/H] $\sim$ -0.8 dex (see Table 1).

\begin{figure}
\includegraphics[width=0.5\textwidth]{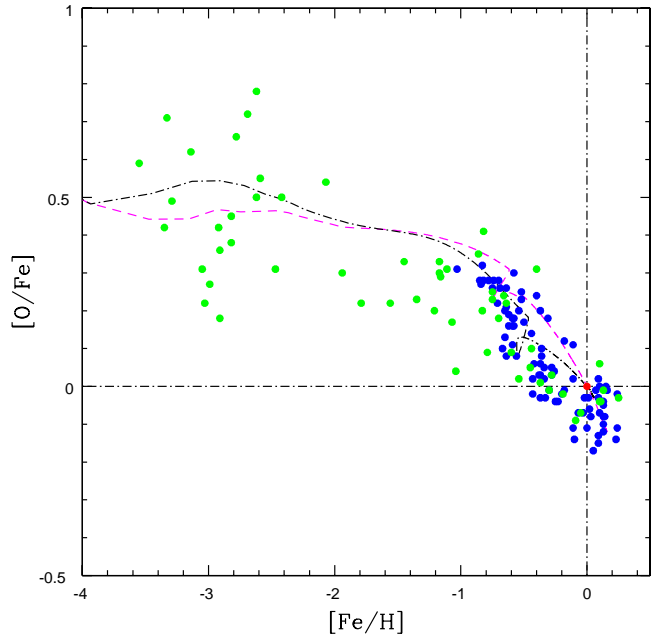}
\caption{The predicted [O/Fe] vs. [Fe/H] in the solar vicinity for two different DTDs: dashed line (DTD of SD model from  MR01), dashed-dotted (DTD of G05 for the wide DD channel). The model results are normalized to their own predicted solar values. The data are taken from the compilation of Fran\c cois et al. (2004).
The solar position is marked by a red dot.}
\label{fig4}
\end{figure}

In Figure 5 we show the predicted [O/Fe] trend for the DTD suggested by S04
where the delay in the appearence of the first Type Ia SNe is much longer than in the 
previous cases. As one can see from the figure, the [O/Fe] ratio in this case shows a 
longer plateau than in the other models, and the knee in the [O/Fe] ratio occurs at [Fe/H]$\sim$ -0.6 dex. 
It is worth noting that the cosmic epoch at which the knee in 
the [O/Fe] ratio occurs in the models with the SD and DD rates is $\sim$1 Gyr, whereas in the case of 
Strolger et al. DTD this time is 4-5 Gyr.
In Figure 5 we also show a comparison between the results obtained with a DTD $\propto t^{-0.5}$ (PHS08) 
and the DTD of MVP06.
As one can see, the PHS08 DTD produces results not very different from the SD model but it predicts a too low solar Fe abundance ($[Fe/H]_{\odot}$=-0.16dex) and a too high solar $[O/Fe]_{\odot}$=0.22dex, as opposed to all the other models where the solar $[Fe/H]_{\odot}$ and $[O/Fe]_{\odot}$ are very close to zero (see Table 1 where we show the predicted solar ratios). Note that the model results are always normalized to their own predicted solar values.

\begin{table*}
\caption{Predicted $[O/Fe]_{\odot}$ and $[Fe/H]_{\odot}$ values at the time of formation of the solar system plus the value of $[Fe/H]_{knee}$ at which the knee for each model occurs.
Normalization to the solar values of Asplund et al. (2005).}
\centering
\begin{tabular}{c|c|c|c|c}
\noalign{\smallskip}
\hline
\hline
\noalign{\smallskip}
Model & $[O/Fe]_{\odot}$ & $[Fe/H]_{\odot}$ & $[Fe/H]_{knee}$ \\
\noalign{\smallskip}
\hline
\noalign{\smallskip}
SD    & 0.095      & -0.013  &  -0.8\\
DD    & 0.095     & -0.038 &  -0.8\\
MVP06 & 0.07      & -0.003  &  -1.8\\
S04   & 0.06      & 0.01 &  -0.6\\
PHS08 & 0.22      & -0.16  &  -0.45\\
\noalign{\smallskip}
\hline
\hline
\end{tabular}
\end{table*}

\begin{figure}
\includegraphics[width=0.5\textwidth]{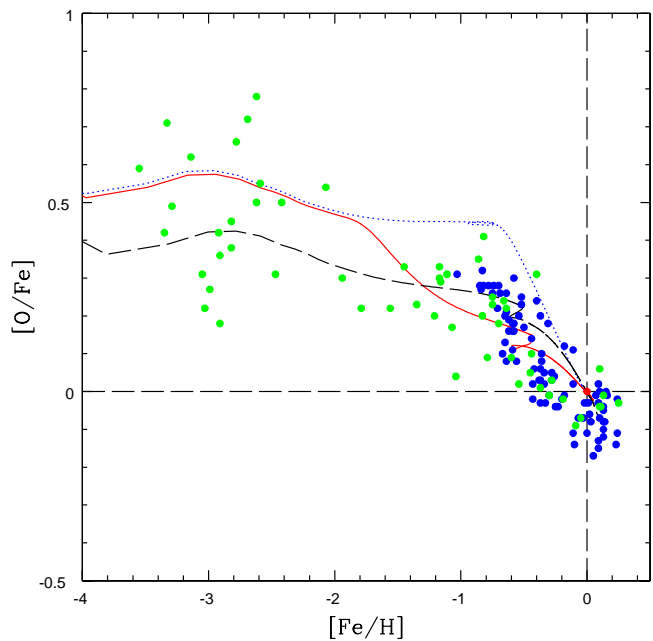}
\caption{In this figure we show the predicted [O/Fe] when the Strolger et al. DTD is assumed (dotted line) 
compared with the predictions from the models with MVP06 DTS (continuous line)  and the model with the  DTD$\propto t^{-0.5}$ (PHS08) (long dashed line. The model results are normalized to their own predicted solar values. The data are the same as in Fig. 4.}
\label{fig5}
\end{figure}

\begin{figure}
\includegraphics[width=0.5\textwidth]{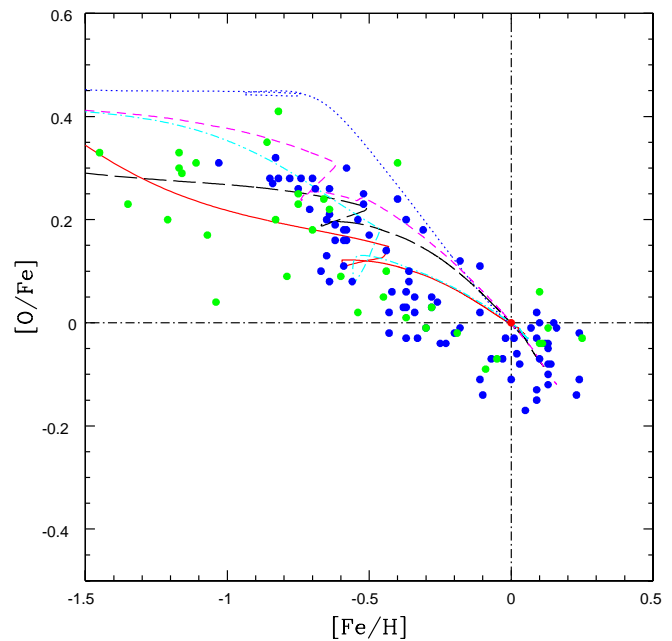}
\caption{Comparison of all the models with all the studied DTDs: the short dashed curve refers to the DTD of MR01; the continuous curve refers tothe DTD of MVP06; the dotted line represents S04 DTD; the long dashed curve refers to the DTD of PHS08; the dashed-dotted curve refers to the DTD of the DD scenario (wide channel) of G05. The model results are normalized to their own predicted solar values. The data are the same as in Fig.4.}
\label{fig6}
\end{figure}


\begin{figure}
\includegraphics[width=0.5\textwidth]{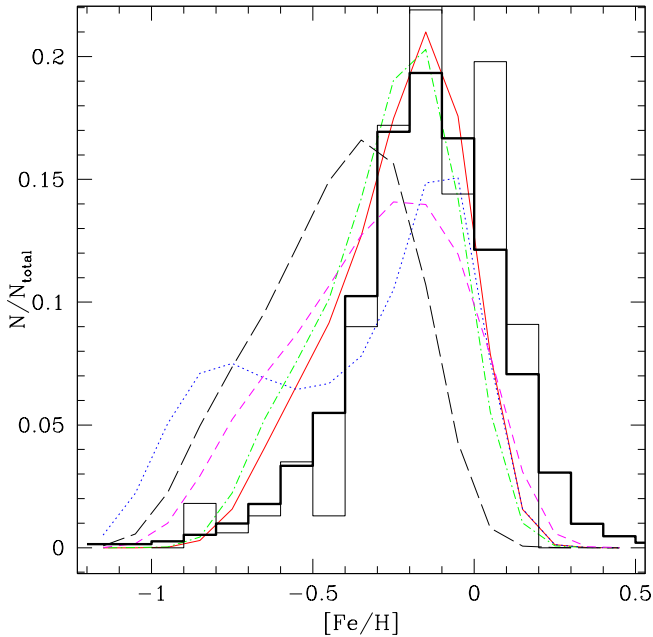}
\caption{The G-dwarf metallicity distribution in the solar vicinity. The data 
are from the Geneva-Copenhagen Survey of the Solar Neighbourhood, Nordstr{\"o}m et al. (2004) 
(thick line histogram) and the data by J{\o}rgensen (2000) (thin-line histogram).
The models refer to: S04 DTD (dotted line), PSH08 DTD (long dashed line), 
DD model (dashed-dotted line), MVP DTD (continuous line) and SD model (short dashed line).}
\label{fig7}
\end{figure}

Finally, in Figure 6 we show a zoomed comparison of all the studied DTDs in the [Fe/H] range -1.5--0.5 dex:
here the different effects produced by the different assumptions about Type Ia SN progenitors are more evident. A larger number of prompt Type Ia SNe clearly produces lower [O/Fe] ratios in this [Fe/H] range with respect to models depressing the number of prompt SNe in favor of tardy ones. Although the spread existing in the data prevents us from drawing firm conclusions, from this comparison it turns out that the S04 model tends to predict too high values of [O/Fe] in the range   -1.5 $\le$ [Fe/H]$\le$ -0.2.On the other hand, the model with MVP06 DTD predicts the lowest [O/Fe] ratios in the range -1.8 $\le$ [Fe/H]$\le$ -1.0. From numerical experiments, we can say that the MVP06 DTD with $\sim30\%$ of prompt SNe would fit  better the data, and it would be still compatible with the findings of MVP06.
It is also interesting to note that the DTDs influence the [Fe/H] at which the gap in the star formation between the halo-thick disk and the thin disk phases is visible. This gap produces a loop in the [O/Fe] behaviour and is naturally produced by assuming a threshold gas density in the star formation. Observationally, a gap in the abundance ratios has been claimed by Fuhrmann (1998) and Gratton et al. (2000). In particular, the gap was observed at a [Fe/H] between -1.0 and -0.5 dex, corresponding to the transition phase between the thick and the thin disk. If confirmed, this gap can impose constraints on the history of star formation.
In our plots the gap occurs at larger [Fe/H] values going from the SD to PHS08, DD and MVP06 DTD (see Figure 6). For the S04 DTD, the gap is not visible and the reason is that Type Ia SNe start exploding in a non negligible number only after the gap. Therefore, the effect of the halt of oxygen production, while the Fe is still produced by the Type Ia SNe exploding during the gap, is lost. This effect produces in the other models the loop in the [O/Fe] ratio. If the gap really exists this would further strenghten the conclusion that prompt Type Ia SNe should exist and be a non negligible fraction of all SNe Ia.\\
In summary, the SD and DD models together with the bimodal distribution of MVP06 all produce acceptable results, although a 50\% of prompt Type Ia SNe appears too large. The DTD from PHS08 and S04 are probably less likely although the spread in the data prevents us from drawing firm conclusions.\\
Therefore, we checked our results on a very important observational constraint: the G-dwarf metallicity distribution.
In Figure 7 we show the predictions for the G-dwarf metallicity distribution of all our models compared with observational data. This Figure shows that two models can be ruled out: the PHS08 and the S04. Both of these models contain empirically derived DTDs. They predict too many 
metal poor stars below [Fe/H]$<$ -1.0 dex (the so-called G-dwarf problem), and this is due to the fact that in both cases the Fe abundance increases more slowly that in the other three models (SD, DD, MVP06) thus producing too many metal poor stars.The reason for that resides in the small number of prompt Type I a SNe in the PHS08 and S04 models. In particular, the S04 model shows two distinct peaks in the G-dwarf metallicity distribution and this is due to the fact that at the beginning of the thin disk formation only SNeII enrich the gas in Fe and the bulk of Fe comes with a large delay, thus creating the second peak. On the other hand, the other models all produce acceptable results, with  the DD and MVP06 models being the best. Note that all the models predict less metal rich stars than observed by Norstr{\"o}m et al. (2004) which is the largest and more accurate survey of G and F dwarfs. This is a problem for pure models of chemical evolution which do not take into account the possibility of stellar migration. In fact, because of interaction between stars and transient spiral density waves (Roskar et al. 2008), stars born in the inner part of the disk can be scattered at larger galactocentric distances and this could be the explanation for the existence of metal rich stars in the solar vicinity.

\section{Conclusions}

We have studied the effects of different DTDs on the [O/Fe] vs.[Fe/H] diagram, where the effects of the time delay in the Fe production by Type Ia SNe are particularly important, and on the G-dwarf -[Fe/H] distribution. We considered the DTDs related to the two main progenitor models for SNe Ia, namely the SD and DD models. Then we considered other DTDs derived empirically from the study of SN rates in galaxies (MVP06, PHS08) and the cosmic SN rate (S04), which are not directly related to particular progenitor models.
We computed the chemical evolution of the solar vicinity by adopting a successful model of Galactic chemical evolution, 
the so-called two-infall model. 
Our main results can be summarized as follows:

\begin{itemize}
\item Delay time distributions containing a not negligible number of prompt Type a SNe (namely those SNe exploding inside 100 Myr since the beginning of star formation) are necessary to best fit the [O/Fe] ratio. This indicates that the first Type Ia SNe occurring in galaxies can explode as soon as after 35-40 Myr since the beginning of star formation. This corresponds to the lifetime of a 8$M_{\odot}$ star which is also the maximum mass for the progenitors of C-O white dwarfs.As a consequence of this, the very first Type Ia SNe already appear during the halo phase.

\item The bimodal DTD of MVP06 predicts a very high fraction of prompt Type Ia SNe ($\sim 50\%$). This fraction is probably too high since it depresses the [O/Fe] ratio in the  [Fe/H] range
-1.8--0.8 dex and produces a knee in the [O/Fe] ratio at a too low metallicy (the observed knee is at [Fe/H] $\sim -1.0$ dex). A smaller fraction of promt SNe Ia would be acceptable (see also  Matteucci et al. 2006).

\item The DTD of S04, which predicts the maximum of Type Ia SNe occurring at 3.4 Gyr since the beginning of star formation, overproduces the [O/Fe] ratio in the [Fe/H] range -1.0--$\,\,$ -0.2 dex and predicts a knee in the  [O/Fe] ratio occurring at a too high [Fe/H] $\sim$ -0.6 dex. Although the spread in the data prevents us from drawing firm conclusions the best DTDs from the point of view of the [O/Fe] diagram are the one related to the SD and DD cases.

\item The G-dwarf metallicity distributions predicted by the models with different DTDs give a more clear indication on the DTDs which should be discarded and those are the two empirical ones derived by S04 and PSH08. They
produce too many disk stars with metallicity [Fe/H]$< $ -1.0 dex. Therefore, we conclude that the DTDs related to the two classical models for Type Ia SN progenitors (SD and DD), and in particular that related to the DD model (wide channel,) are the best to reproduce the observational data.
Since these two models include prompt Type Ia SNe even if in a smaller proportion than the bimodal DTD 
of MVP06, we conclude again that such prompt SNe are necessary to reproduce the chemical evolution of the solar vicinity. Probably a mixed scenario (SD+DD) would be the most realistic one
(see Greggio et al. 2008). 

\item On the basis of the results discussed above we can identify the typical timescale for Fe enrichment by Type Ia SNe in the solar vicinity, identified by the time at which the knee in the [O/Fe] ratio occurs, as  1-1.5 Gyr.

\end{itemize}

\section{Acknowledgments}
This work was partially supported by the Italian Space Agency
through contract ASI-INAF I/016/07/0.
F.M., E.S. and S.R. acknowledge financial support from 
PRIN2007-MUR (Italian Ministry of University and Research), 
Prot.2007JJC53X\_001. We also thanks an anonymous referee for his/her useful 
comments.S.R. acknowledges financial support from the FWF through the Lise Meitner grant M1079-N16.

\label{lastpage}

\end{document}